\documentclass[twocolumn,english,prx,showpacs,longbibliography]{revtex4-1}
\usepackage{mathrsfs}
\usepackage{epsfig}
\usepackage{graphicx}
\usepackage{amsfonts}
\usepackage[figuresright]{rotating}
\usepackage{amssymb}
\usepackage{amsmath}
\usepackage{dcolumn}
\usepackage{bm}
\usepackage{color}

\def\be{\begin{equation}} \def\ee{\end{equation}}
\def\bea{\begin{eqnarray}} \def\eea{\end{eqnarray}}

\newcommand{\WQC} {Wilczek Quantum Center and Key Laboratory of Artificial Structures and Quantum Control, School of Physics and Astronomy, Shanghai Jiao Tong University, Shanghai 200240, China}

\newcommand{\SRCQC}{Shanghai Research Center for Quantum Sciences, Shanghai 201315, China}

\begin{document}
\title{Spontaneous symmetry breaking and localization in nonequilibrium steady states of interactive quantum systems}

\author{Shuohang Wu }
\affiliation{\WQC}


\author{Zi Cai}
\email{zcai@sjtu.edu.cn}
\affiliation{\WQC}
\affiliation{\SRCQC}

\begin{abstract}  

The time evolution of a physical system is generally described by a differential equation, which can be solved numerically by adopting a difference scheme with space-time discretization. This discretization, as a numerical artifact, results in accumulated errors during evolution thus usually plays a negative role in simulations. In a quantum circuit, however, the ``evolution time'' is represented by the  depth of the circuit layer, thus is intrinsically discrete. Hence, the discretization-induced error therein is not a numerical artifact, but a physical observable effect responsible for remarkable nonequilibrium phenomena absent in conventional quantum dynamics. In this paper, we show that the combination of measurement feedback and temporal discretization can give rise to a new type of quantum dynamics characterized by unitary but dissipative evolution. As physical consequences of  such an unitary but dissipative evolution, a nonequilibrium steady state with spontaneous symmetry breaking is revealed in a zero-dimensional (single-qubit) system. A localization mechanism distinct from that in the well-established Anderson localization has also been proposed in an one-dimensional interactive quantum system.


\end{abstract}


\maketitle

{\it Introduction --} Introducing quantum measurement and feedback into quantum many-body physics has brought  new opportunities to this field, and attracted considerable interests from both the quantum information and condensed matter communities. For example, the competition between random measurements and unitary dynamics  could give rise to an entanglement transition absent in traditional condensed matter materials~\cite{Skinner2019,Li2018,Li2019,Cao2019,Bao2020,Jian2020,Gullans2020,Chen2020b,Zabalo2020,Turkeshi2020,Lavasani2021,Buchhold2021,Ippoliti2021}. If the measurement outcome could be fed back into the systems and affect on subsequential evolution, the resulting interactive quantum dynamics does not only play an important role in quantum optimization~\cite{Magann2022}, cooling~\cite{Yamaguchi2022},  error correction~\cite{Terhal2015} and  state preparation~\cite{Hurst2020,Wu2022,Lee2022}, but also opens up new avenues to explore the nonequilibrium quantum phenomena such as feedback-induced nonlinear quantum dynamics~\cite{Lloyd2000,Munoz2020}, quantum dynamics with adaptive Trotter time steps~\cite{Zhao2023} and  nonequilibrium phases~\cite{McGinley2022} and phase transitions~\cite{Ivanov2020}. Experimentally, this interactive quantum dynamics can be realized in quantum simulators whose  parameters can be dynamically modified according to the measurement outcome.

A target of quantum simulation is to explore the complex ground states of interacting quantum systems from the quantum material,  high-energy physics and quantum chemistry. To approach the ground state, one needs to cool down the system, which usually can be achieved by coupling the system of interest to a low-temperature bath where the excess energy of the system flows to. However, coupling a quantum system to a bath typically breaks the unitarity of its evolution, and the bath brings not only dissipation, but also decoherence, which will result in a mixed state rather than a pure ground state. To avoid this side effect of bath, one may wonder whether it is possible to realize a quantum dynamics during which the system energy is dissipated while the unitarity is kept.

In this study, we propose such an interactive quantum dynamics in an one-dimensional (1D) quantum lattice model, where the local density is repeatedly measured at a sequence of time slices with an equal interval $T$. After each measurement, the outcome is fed back to the system by modifying its onsite chemical potential according to the measurement outcome.  This model exhibits a dissipative but unitary evolution driving  the system into nonequilibrium steady states, which could exhibit intriguing phenomena such as spontaneous symmetry breaking (SSB) in a zero-dimensional system, and localization with a mechanism distinct from the well-established Anderson localization~\cite{Anderson1958}.

 \begin{figure}[htb]
\includegraphics[width=0.99\linewidth]{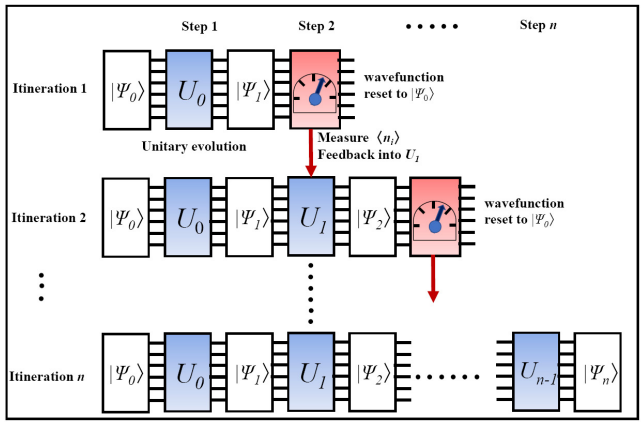}
\caption{(Color online) Schematic showing procedure for implementing the interactive dynamics. Starting from an initial state $|\psi_0 \rangle$,  in itineration 1, we first operate $\hat{U}_0$ on $|\psi_0 \rangle$ to derive $|\psi_1\rangle$, then  measure its average local density and prepare $\hat{U}_1$ for next itineration. Finally, we reset the wavefunction to $|\psi_0 \rangle$, and start itineration 2 following a similar procedure (deriving $|\psi_2\rangle$, measuring $\langle \hat{n}_i\rangle$ and preparing $\hat{U}_2$,  resetting the wavefunction to $|\psi_0\rangle$).  We repeat this procedure until itineration n, at the end of which we realize a wavefunction $|\psi_n \rangle=\hat{U}_{n-1}\cdots \hat{U}_0 |\psi_0\rangle$. }\label{fig:fig1}
\end{figure}

{\it The model and its physical realization--} We consider a quantum dynamics described by a sequence of unitary evolutions:
\begin{equation}
|\psi_n\rangle= \hat{U}_{n-1}\cdots \hat{U}_0 |\psi_0\rangle\label{eq:evolution}.
\end{equation}
where $|\psi_0\rangle$ is the initial state wavefunction at $t_0=0$  and $|\psi_n\rangle$ is the wavefunction at the $n$-th time slice with $t_n=nT$.
$\hat{U}_n=e^{-iT \hat{H}_n}$ is the unitary evolution operator and $\hat{H}_n$ is the Hamiltonian which is invariant during the interval $t\in [t_n,t_{n+1}]$:
\begin{equation}
\hat{H}_n=\sum_i [-J(\hat{c}_i^\dag \hat{c}_{i+1}+h.c)-\mu_i^n \hat{n}_i] \label{eq:Ham}
\end{equation}
where $i$ and $n$ denote the site and time step index respectively. $\hat{c}_i$ ($\hat{c}_i^\dag$) is the annihilation (creation) operator of the spinless fermion on site $i$ and $\hat{n}_i=\hat{c}_i^\dag \hat{c}_i$. $J$ is the nearest-neighboring single-particle hopping amplitude and $\mu_i^n$ is the onsite chemical potential which is invariant during the interval $t\in [t_n,t_{n+1}]$. The feedback is introduced in a way that $\mu_i^n$ in $\hat{H}_n$ is determined by the wavefunction $|\psi_{n}\rangle$ at the time slice ($t=t_{n}$)  as:
\begin{equation}
\mu_i^n=\mu_i+V \langle\psi_{n}|\hat{n}_i|\psi_{n}\rangle \label{eq:chemical}
\end{equation}
where  $\mu_i$ is randomly sampled from a uniform random distribution with $\mu_i\in[-\Delta,\Delta]$ and $\mu_i^0=\mu_i$.  The parameter $V$ characterizes the strength of feedback. The sign of $V$  is important, throughout the paper, we choose $V>0$ which indicates an positive feedback mechanism: once a particle is found to be localized on site $i$, the on-site potential is further lowered in the subsequential evolution, which, in turn, enhances the tendency of localization. This positive feedback mechanism is crucial for our discussion in the following.

To collect the information of local density, one needs to perform measurement after each time step. However, these measurements will lead to wavefunction collapse, which has not been taken into account in the evolution defined in Eq.(\ref{eq:evolution}). To preserve the unitarity of the evolution and collect the information of the physical quantities in the meantime, we use an  iterative feedback-based algorithm for quantum circuit\cite{Magann2022} as illustrated in Fig.\ref{fig:fig1}. The basic idea is that after the $n$-th time step, we measure the physical quantities over the wavefunction $|\psi_n\rangle$ and prepare the evolution operator $\hat{U}_{n}$, then reset the wavefunction to the $|\psi_0\rangle$ and start over from beginning until it reaches the $n$-th time step. We then operate the preliminarily prepared operator $\hat{U}_{n}$ on $|\psi_n\rangle$ to obtain $|\psi_{n+1}\rangle$. Consequently, to obtain the wavefunction $|\psi_n\rangle$, one has to perform  $\frac{n(n+1)}2$ unitary operations  in total. In the following, we will analyze the consequence of this feedback-induced interactive dynamics.

{\it The simplest albeit nontrivial case with $L=2$ --} We start with the simplest case of one particle in a two-site lattice. Despite its extreme simplicity, this model captures the most important ingredients of  interactive dynamics: feedback-induced nonlinearity and dissipation. This simple model, with a Hilbert space dimension of 2, can be mapped to a single spin-$\frac 12$ model ($|10\rangle\rightarrow |\uparrow\rangle$ and $|01\rangle\rightarrow |\downarrow\rangle$)  with a time-dependent magnetic field whose strength and direction are, in turn, determined by the spin itself, where the Hamiltonian.(\ref{eq:Ham}) turns to:
\begin{equation}
\hat{H}_n=-[2V s^z_{n}+2\Delta]\hat{s}^z- 2J \hat{s}^x\label{eq:Hamspin2},
\end{equation}
where $s^z_{n}=\langle\psi_{n}|\hat{s}^z|\psi_{n}\rangle$ is a step-like function, $\hat{s}^\alpha=\frac 12 \sigma^\alpha$ with $\sigma^\alpha$ being the Pauli matrices, and $\Delta=\mu_1-\mu_2$.  In the continuous time limit ($T\rightarrow 0$), the equation of motion (EOM) of this model become a nonlinear differential equation taking the same form as the mean-field  EOM of the Lipkin-Meshkov-Glick(LMG) model~\cite{Lipkin1965}. However, there is crucial difference: the LMG model is an infinite-dimensional model while our model is the zero-dimensional (single spin model), where the physical origin of the nonlinearity  is  feedback rather than interaction. Typically, the LMG model  exhibits persistent oscillation, which depends strongly on the initial state. However, the intrinsic temporal discreteness (finite T) in our interactive dynamics will introduce dissipation and qualitatively alter the long-time dynamics.

\begin{figure}[htb]
\includegraphics[width=0.99\linewidth,bb=170 58 1150 550]{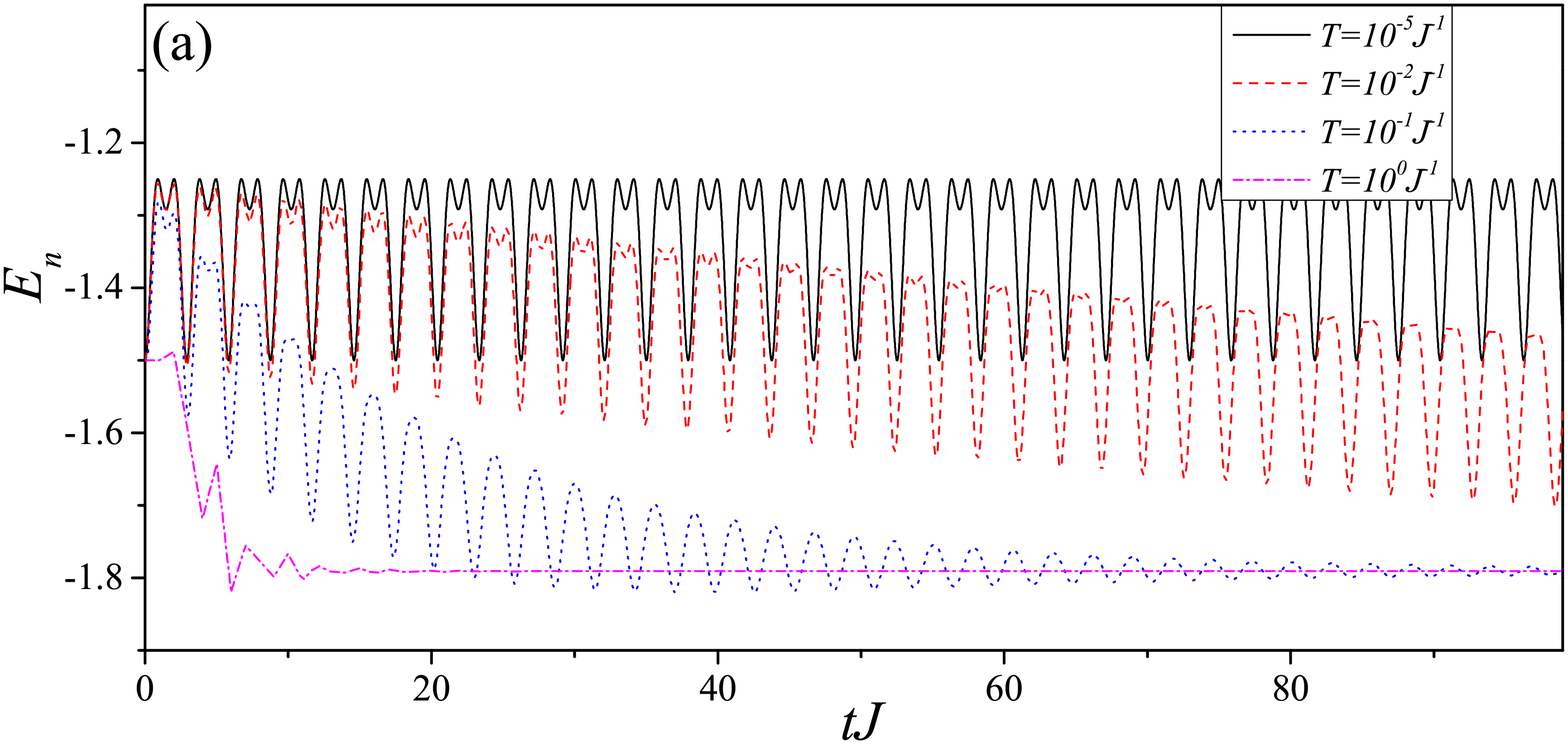}
\includegraphics[width=0.99\linewidth,bb=130 55 1220 555]{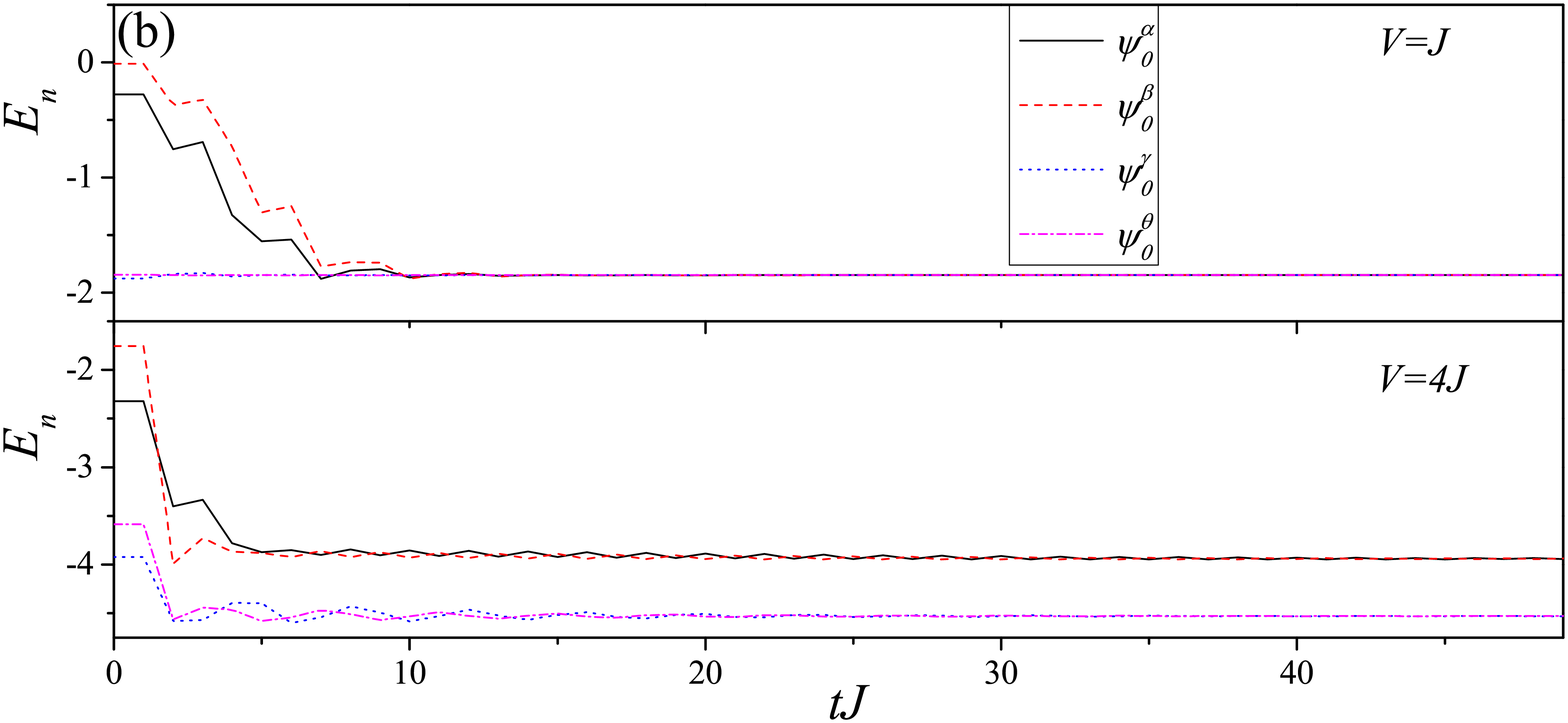}
\caption{(Color online) (a) Dynamics of the instantaneous energy $E_n=\langle\psi_n|\hat{H}_n|\psi_n\rangle$ starting from the initial state $|\psi_0\rangle=|\uparrow\rangle$ for the single-spin system (Hamiltonian.\ref{eq:Hamspin2}) with different time intervals (T). The parameters are chosen as $V=J$, $\Delta=J$.  (b) Dynamics of $E_n$ starting from different initial states in the presence of a small $V=J$ with only one steady state (upper panel) and a large $V=4J$ with two steady states (lower panel).  The initial states are chosen as $|\psi_0^{\alpha/\gamma}\rangle=\sqrt{0.8}|\uparrow\rangle\pm\sqrt{0.2}|\downarrow\rangle$ and $|\psi_0^{\beta/\theta}\rangle=\sqrt{0.7}|\uparrow\rangle\pm\sqrt{0.3}|\downarrow\rangle$.   Other parameters are chosen as $T=J^{-1}$ and $\Delta=0.5J$.
} \label{fig:fig2}
\end{figure}

The  instantaneous energy $E_n=\langle \psi_n|H_n|\psi_n\rangle$ as a function of evolution time $t=nT$ for different $T$, comprising  a fixed $V$ has been plotted in  Fig.\ref{fig:fig2} (a). After sufficiently long time,  $E_n$ keeps being dissipated for a finite T and approaches a minimum value corresponding to the ground state energy of the mean-field LMG model (up to a constant). This energy dissipation is not due to the coupling to a bath, since the evolution in Eq.(\ref{eq:evolution}) is unitary. Instead, it is a reminiscence of the artificial dissipation induced by temporal discretization in the classical simulation of Hamiltonian dynamics, where the Liouville theorem can be violated if we choose a dissipative difference scheme.  However, this numerical artifact strongly depends on the difference scheme, thus can be eliminated by choosing other difference schemes such as a symplectic geometric algorithm~\cite{Feng1985}. Different from the classical simulation, the time evolution is intrinsically discrete in our quantum model, where the ``difference scheme''  has already been fixed by the iterative feedback-based procedure as shown in Fig.\ref{fig:fig1}.

 Subsequently, we fix $T$ and check the dependence of long-time dynamics on the feedback parameter $V$. As shown in Fig.\ref{fig:fig2} (b), for a small $V$, the long-time evolutions always converge to the same steady state irrespective of the initial states. However, when $V$ exceeds a threshhold, a new steady state with higher energy emerges, and the system could fall into either steady state depending on the initial state. It is worth emphasizing that these two steady states are not orthogonal with each other, and an arbitrary superposition of them is no longer steady due to the nonlinearity of the EOM. If $V$ is further increased, more steady states will emerge. These steady states act as attracting basins for different initial states, and the physical meaning of these steady states can be understood as a consequence of the temporal discretization. These states are ``steady'' in the sense that they are robust against small perturbations imposed on them. However, a strong perturbation may drive the system out of the original basin, thus these steady states are actually ``metastable''.

\begin{figure}[htb]
\includegraphics[width=0.99\linewidth,bb=170 56 1200 555]{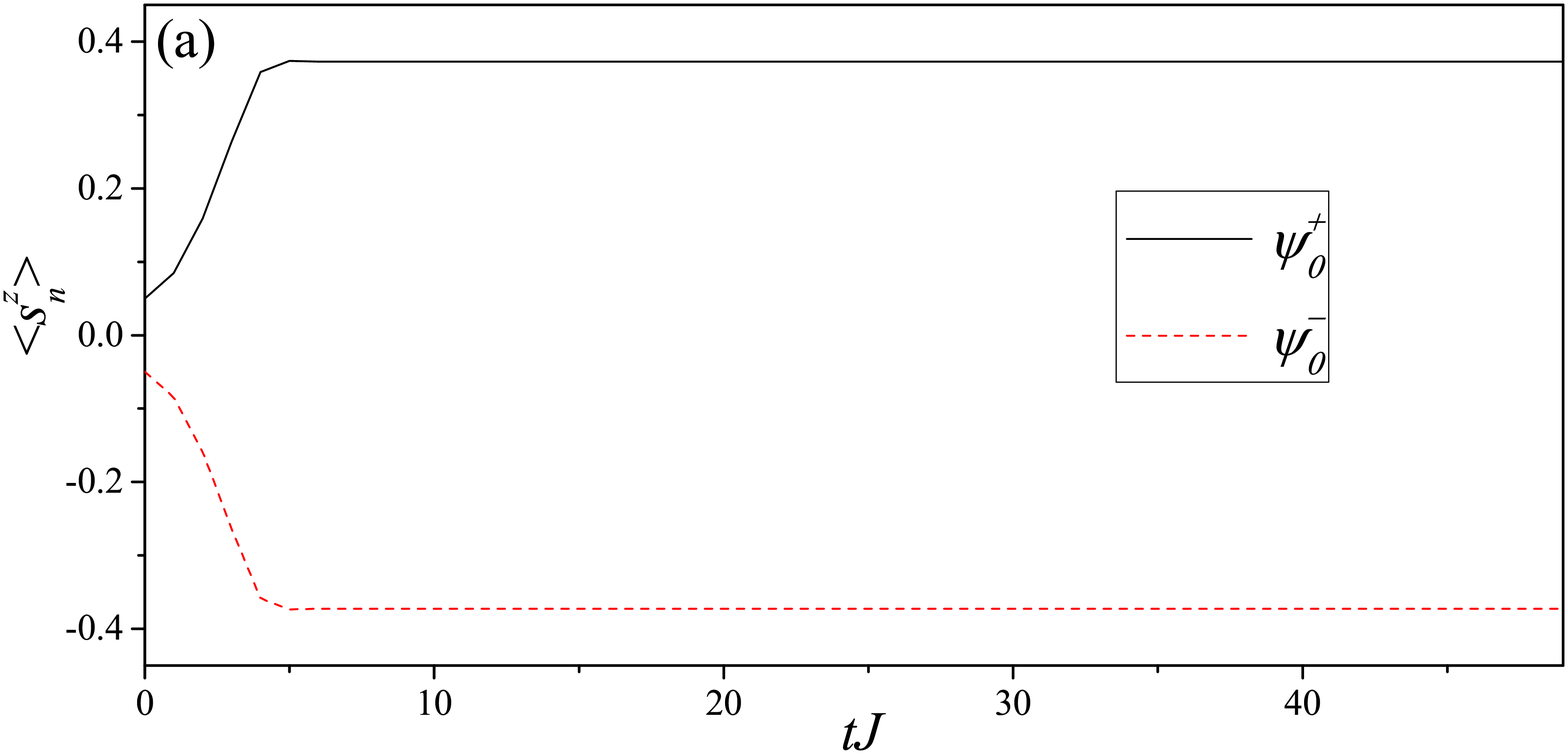}
\includegraphics[width=0.99\linewidth,bb=170 56 1200 555]{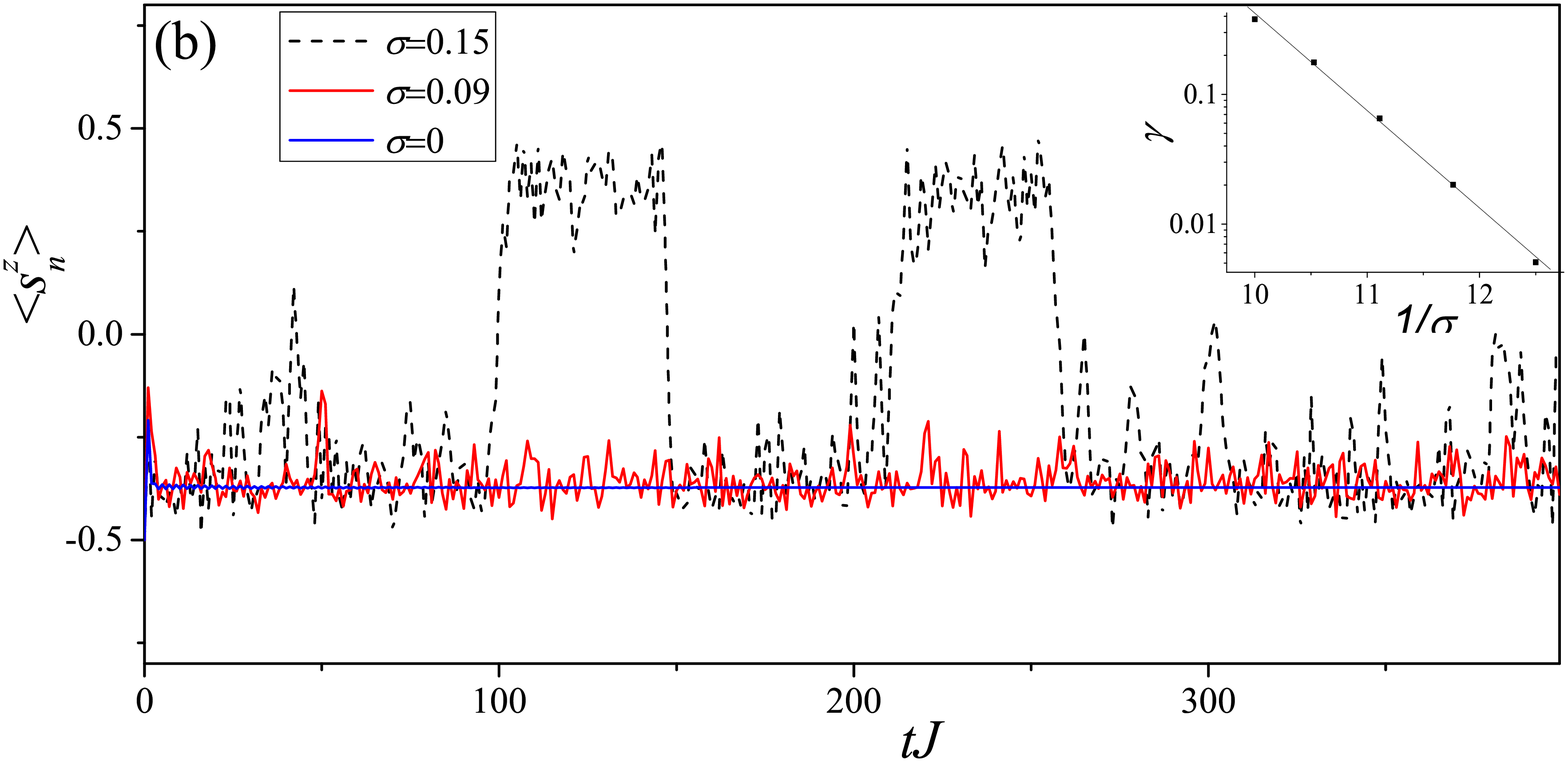}
\caption{(Color online)  (a) Dynamics of  $\langle s_n^z\rangle$ for the single spin system without bias ($\Delta=0$ in Hamiltonian.\ref{eq:Hamspin2}) starting from different initial states $|\psi_0^+\rangle=\sqrt{0.55}|\uparrow\rangle+\sqrt{0.45}|\downarrow\rangle$ and $|\psi_0^-\rangle=\sqrt{0.45}|\uparrow\rangle+\sqrt{0.55}|\downarrow\rangle$.   (b) Dynamics of $\langle s_n^z\rangle$ starting from an initial state $|\psi_0\rangle=|\downarrow\rangle$ under single noise trajectory with difference noise strength $\sigma$ (The inset indicates the tunneling rate $\gamma$ as a function of $\frac 1\sigma$  after ensemble average over $10^2$ noise trajectories).  The parameters are chosen as $V=3J$, $\Delta=0$ and $T=J^{-1}$.
} \label{fig:noise}
\end{figure}

{\it Spontaneous symmetry breaking in a zero-dimensional non-equilibrium steady states--} More interesting phenomenon occurs in the  $L=2$ case without bias ($\Delta=0$), where the single-spin system preserves the $Z_2$ symmetry: Hamiltonian.(\ref{eq:Hamspin2}) is invariant under the transformation $s^z\rightarrow -s^z$, which corresponds to the inverse symmetry $(i=1\leftrightarrow 2)$ in the original Hamiltonian.(\ref{eq:Ham}). However, as shown in Fig.\ref{fig:noise} (a), for $V>2J$ the steady state spontaneously breaks this $Z_2$ symmetry with an  order parameter $\langle \hat{s}^z\rangle\neq 0$ depending on the initial state.  Although the non-equilibrium steady states in our model share the same mathematic structure with the mean-field ground state of the  LMG model, there are important difference. The LMG model is defined in infinite dimension, while our single-spin Hamiltonian.(\ref{eq:Hamspin2}) provides an example of zero-dimensional SSB absent in any thermal equilibrium system, thus is a genuine nonequilibrium phenomenon.  This 0D SSB also reminds us of the  localization transition in the spin-boson model, where strong dissipation can suppress the quantum tunneling  and give rise to a localized phase with $Z_2$ SSB\cite{Leggett1987}. However, the spin-boson model describes a single spin coupled to infinite bath degrees of freedom, thus is not an exact 0D system, while no bath is present in our model.  In our model, the SSB originates from the positive feedback mechanism: once a spin exhibits a polarized  tendency along z direction, this tendency will be enhanced in the subsequential evolution via the feedback mechanism.   This feedback-induced SSB mechanism is distinct from those induced by interaction as in the LMG model or dissipation as in the spin-boson model.

Different from conventional quantum dynamics monitored by repeated measurements, the interactive dynamics in our model is  deterministic  since the feedback quantity is its average value over a set of copies of identical quantum states, instead of an outcome of a single projective measurement, thus the evolution defined in Eq.(\ref{eq:evolution}) does not involve any randomness in principle. However, in realistic quantum systems, the accuracy of an average value of a physical quantity depends on the number of the identical copies prepared in the experiments. The inaccuracy of the measurement outcome will introduce randomness, which can be modeled phenomenologically by include a Gaussian noise in the feedback quantities: $s_n^z$ in Eq.(\ref{eq:Hamspin2}) is replaced by $\tilde{s}_n^z=\delta_n$ with $\delta_n$ being a  stochastic variable satisfying the Gaussian distribution with average value $\langle \delta_n\rangle=s_n$ and variance $\sigma$, which measures the uncertainty of the measurement outcome, and depends on the number of identical copies $\mathcal{N}$ in the measurement as $\sigma\sim \frac 1{\sqrt \mathcal{N}}$.

As shown in Fig.\ref{fig:noise} (b), a finite $\sigma$ will induce a tunneling between the different steady states. However, the tunneling rate $\gamma$ (the average probability of the tunneling from per unit time) is exponentially suppressed in the presence of small $\sigma$. The inset of Fig.\ref{fig:noise} (b) indicates that $\gamma\sim e^{-\frac c\sigma}$, where $c$ is a constant. Considering the fact that $\sigma\sim \frac 1{\sqrt \mathcal{N}}$, this result indicates that in a realistic experimental system, the life-time of a SSB phase $\tau_c\sim \frac 1\gamma$ grows exponentially with $\sqrt{\mathcal{N}}$ ($\tau_c\sim e^{c\sqrt{N}}$), which enables us to observed this SSB phase experimentally provided the the number of identical copies of the wavefunction in the measurement is sufficiently large.

One can introduce other type of noise, for example the decoherence which is inevitable in a realistic
quantum computer, and we believe the result is similar. For a zero-dimensional system, the noise will always
induce a finite-life time for the symmetry breaking phase. However, for a higher dimensional interacting quantum
system, there is a competition between the feedback-induced dissipation and noise-induced heating, which may
result in a non-equilibrium phase transition which resembles the thermal phase transition in equilibrium system.


\begin{figure}[htb]
\includegraphics[width=0.9\linewidth,bb=120 53 900 555]{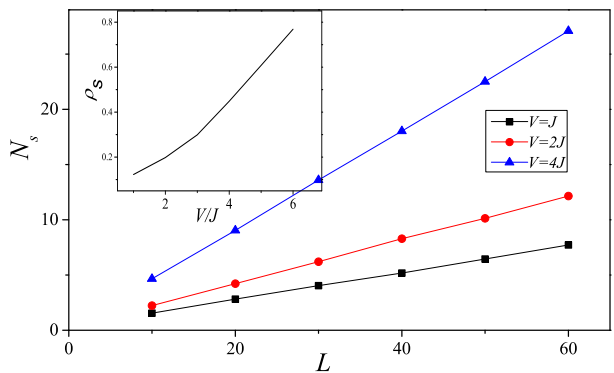}
\caption{(Color online) In a 1D lattice with length $L$, the total number of the steady state $N_s$ as a function of $L$ with different $V$, which exhibit a linear relation $N_s=\rho_s L$.  The inset is the dependence of the steady state density $\rho_s$ on $V$.  $\Delta=2J$.
} \label{fig:fig3}
\end{figure}

\begin{figure*}[htb]
\includegraphics[width=0.325\linewidth,bb=95 53 760 555]{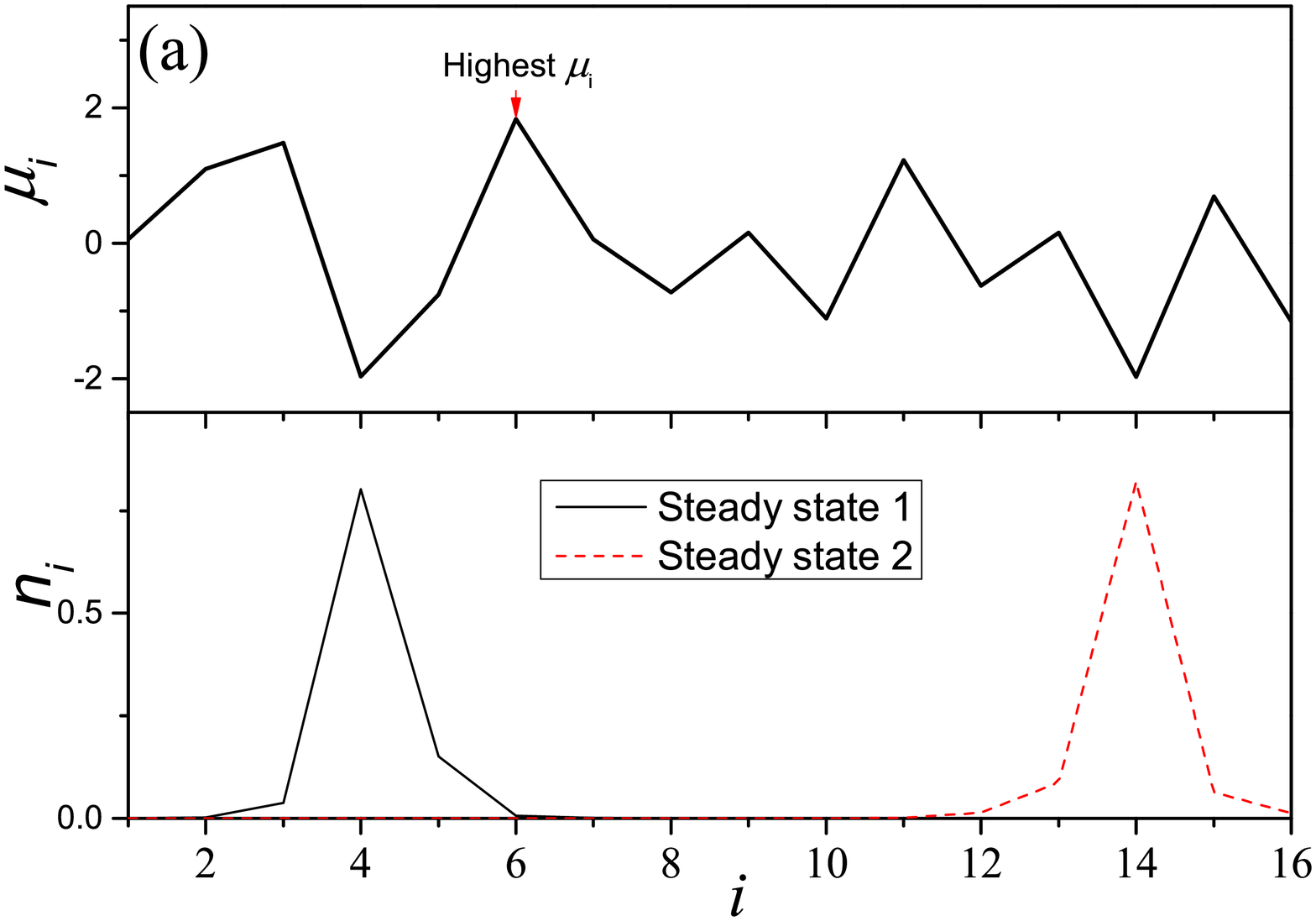}
\includegraphics[width=0.325\linewidth,bb=95 53 760 555]{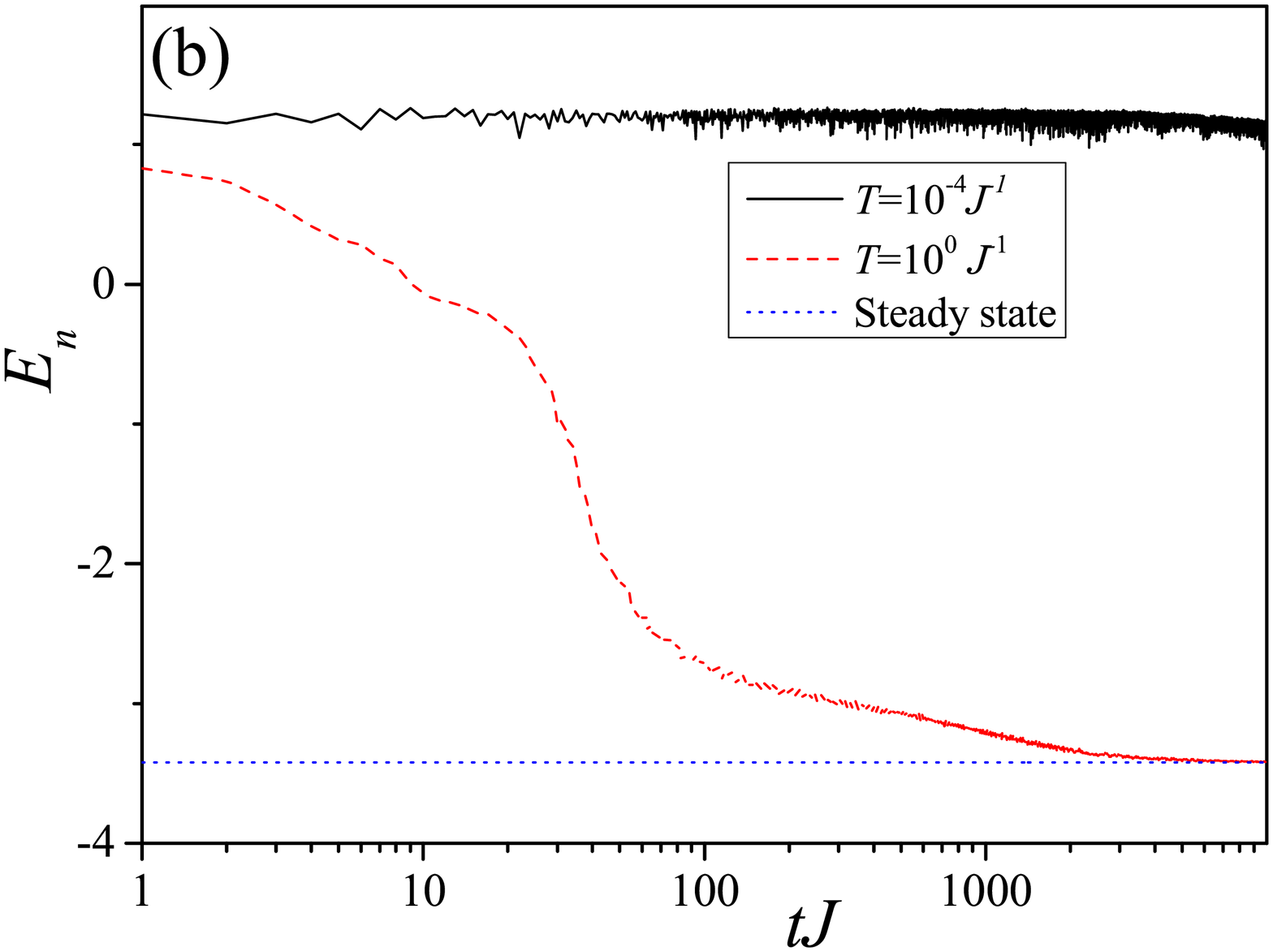}
\includegraphics[width=0.325\linewidth,bb=95 53 760 555]{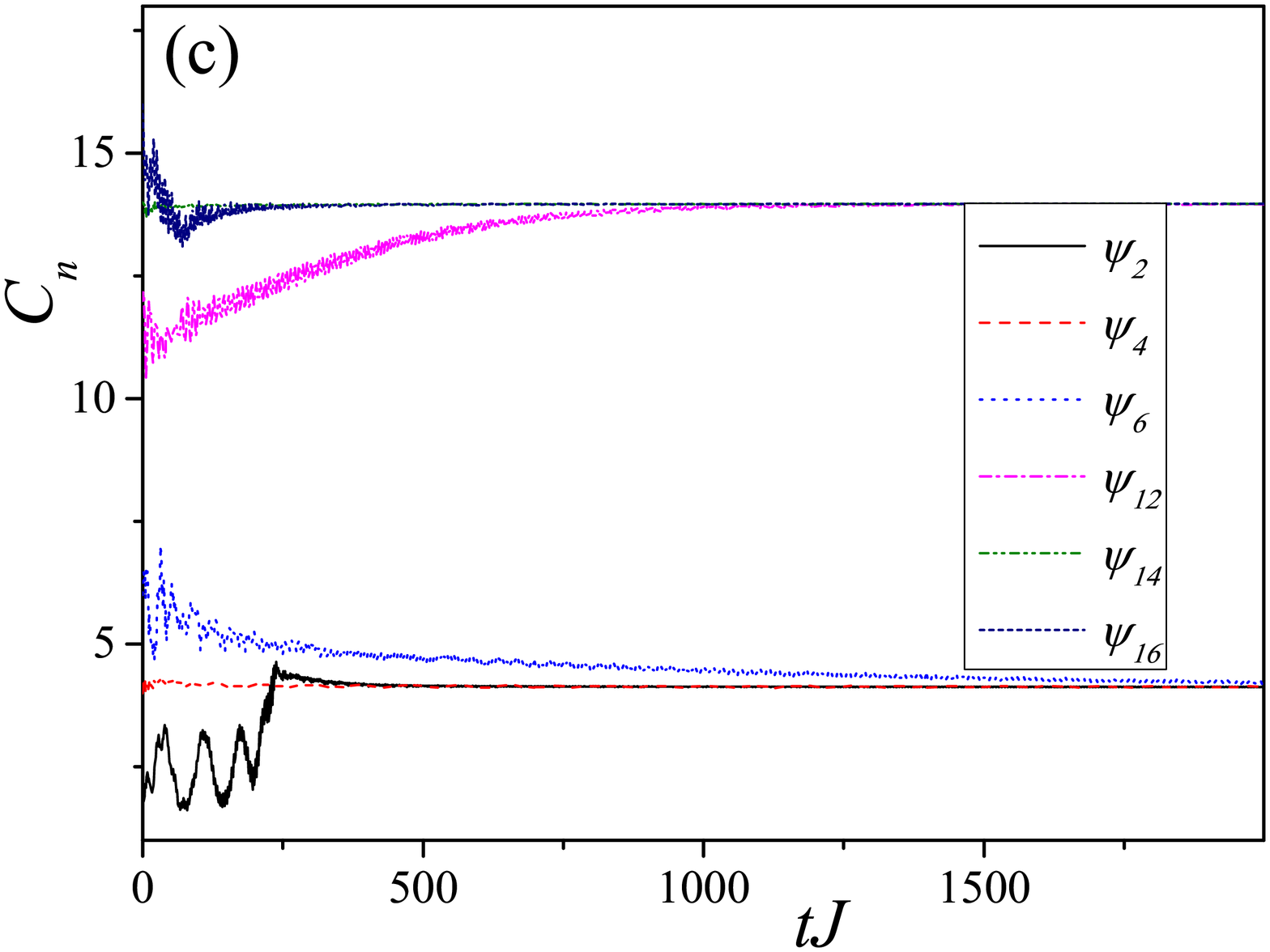}
\caption{(Color online) (a) An $L=16$ 1D lattice with a set of random $\mu_i\in [-\Delta,\Delta]$ (upper panel) and  the corresponding density distribution of the two single-particle steady states(lower panel), each of which is centered at lattice sites ($i=4$ and $14$) with the local minimum potential. (b) Dynamics of $E_n$ starting from an initial state located on the lattice site ($i=6$) with the maximum potential with different T. In the long-time limit, $E_n$ will approach the energy of the steady state located on lattice site $i=4$ (blue dashed line) (c)Dynamics of the center of mass $C_n$ starting from different localized initial states ($\psi_i$ indicates an initial state that the particle is located on site i).  Depending on  its initial position, the particle will finally fall into either steady state 1 or 2. Other parameters are chosen as $\Delta=2J$, $V=J$ for (a)-(c) and $T=J^{-1}$ for (c).
} \label{fig:fig4}
\end{figure*}

{\it Single particle dynamics --} Next, we generalize our discussions to the single-particle dynamics in a 1D lattice in the presence a static disordered  and a dynamically modified chemical potential defined in Eq.(\ref{eq:chemical}). Using an iterative method combined with a stability analysis, we can enumerate all the steady states in such a 1D system. The wavefunctions of these single-particle steady states are spatially local, and their total number is proportional to the system size $N_s=\rho_s L$ as shown in Fig.\ref{fig:fig3}.  For a fixed T, the dependence of the steady state density $\rho_s$  on the feedback factor $V$ is plotted in the inset of Fig.\ref{fig:fig3}. In the following, we will analyze the physical consequences of these steady states.

 In the continuous time limit ($T\rightarrow 0$), the single-particle dynamics is described by a nonlinear differential equation, which takes the same form of a Gross-Pitaevskii (GP) equation~\cite{Pitaevskii2003}, even though the nonlinear term therein has a different origin.  The diffusion dynamics of a wavepacket governed by a GP equation in a 1D disordered lattice has been intensively studied~\cite{Pikovsky2008,Shapiro2007}.  Similar tothe $L=2$ case, the temporal discretization with a finite T qualitatively changes the long-time dynamics of the continuous time limit. To illustrate this point, we consider an example of a small 1D system with only two steady states as shown in Fig.\ref{fig:fig4} (a), and put one particle on the site with the highest $\mu_i$ (say $i=6$), then study its dynamics starting from this initial state. The dynamics of the instantaneous energy $E_n$ with different T is plotted in  Fig.\ref{fig:fig4} (b), from which we can find that for a small T, $E_n$ keeps oscillating around its original value for a long time before starting to decay after a sufficiently long-time (the oscillation become persistent in the continuous time limit). For a larger T, $E_n$ is quickly dissipated and approaches the energy of one of the steady states (steady state 1 centered around site 4).

The dynamics of the wavepacket can be seen more clearly by monitoring its center of mass (COM) defined as $C_n=\sum_i i \langle\psi_n|\hat{n}_i|\psi_n\rangle$. The dependence of the COM dynamics on the initial state is shown in Fig.\ref{fig:fig4} (c), from which we can see that if a particle is injected into the system, it will be attracted and finally trapped by its nearby localized steady state, making it unable to propagate far away from its initial position. For a large $V$ case where the steady state density $\rho_s\simeq 1$, almost every lattice site is centered by a localized steady state wavefunction, and a particle injected into the system will be localized around its original lattice site.   As a result, the information of the initial position of the particle can be (partially) preserved in this interactive dynamics.

It is worthwhile to compare these phenomena with the well-established Anderson localization. Even though they share a lot of common properties such as the spatially localized wavefunction and the preservation of the initial state information, there  are fundamental differences.  The Anderson localization originates from the interference effect of wave, while the localization in our model is due to the extensive metastable states, each acting as a attracting basin that traps nearby particles. This mechanism reminds us of the associative memory in the classical Hopfield model with a set of stored memory patterns~\cite{Hopfield1982}: when a given configuration is similar enough to one of them, the system can retrieve the correct stored pattern via a dissipative dynamics of classical annealing~\cite{Amit1985a,Amit1985b}. Another  difference is that the energy is conserved in the Anderson localization, thus there is no qualitative difference between the localized states with the lowest and highest energies. This similarity, however, does not hold in  in our model due to its intrinsic dissipative nature, where the localized steady states are usually at low energy. As a physical consequence, if a particle is initially located on a lattice site with the highest potential energy, in an Anderson localization system, it will be localized therein. However, in our interactive dynamics,  it will escape from its original site and finally be trapped by a steady state on its nearby site with lower potential energy.

\begin{figure}[htb]
\includegraphics[width=0.9\linewidth,bb=128 62 900 555]{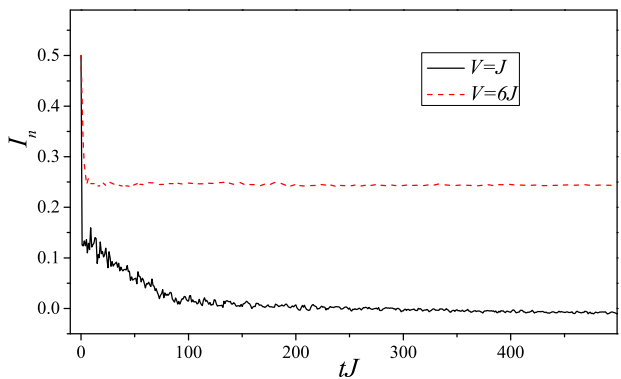}
\caption{(Color online) Dynamics of the initial state memory quantity $I_n$ with different $V$  in a half-filled many-body system. $T=J^{-1}$, $\Delta=2$ and $L=100$. The initial state is chosen as a random product state.
} \label{fig:fig5}
\end{figure}

{\it Many-body dynamics --} Finally, we study the many-body dynamics in our model by considering the case with half-filling. To assess the memory of the initial state information, we choose the initial state as a random product state $|\psi_0\rangle=|n_1 n_2\cdots n_L\rangle$ with $n_i$ being randomly set as 0 or 1, which satisfies $\sum_i n_i=\frac L2$ (half-filling). After an n-step evolution defined in Eq.(\ref{eq:evolution}), the wavefunction turns out to be $|\psi_n\rangle$, from which we could then derive the average value of the local density $\tilde{n}_i=\langle \psi_n|\hat{n}_i|\psi_n\rangle$.  To measure the difference between the density distributions of the initial and final states, we defined a quantity $I_n=\frac 1L\sum_i (-1)^{n_i+1} \tilde{n}_i$, which will reach its maximum value of $\frac 12$ if $\tilde{n}_i=n_i$ $\forall$ i, and will approach zero in the thermodynamic limit if $\{\tilde{n}_i\}$ and $\{ n_i\}$ are uncorrelated. Notice that for a special initial state $|\psi_0\rangle=|1010 \cdots 10\rangle$,   $I_n=\frac 1L\sum_i (-1)^{i+1} \tilde{n}_i$ is the density imbalance between the even and odd sublattice, which has been widely used as an experimental signal of many-body localization~\cite{Schreiber2015}. As shown in Fig.\ref{fig:fig5}, for a large $V$, $I_n$ converges to a finite value, indicating a memory of  the position information of the initial state after a sufficiently long time. For a small $V$ case with a low $\rho_s$, $I_n$ decays to a small value close to zero, indicating that the position information of the initial state has been washed out during the evolution. We expect a crossover instead of a phase transition occurs between the small and large $V$ cases, since $\rho_s(V)$ is an analytic  function, as shown in the inset of Fig.\ref{fig:fig3}.

\begin{figure}[htb]
\includegraphics[width=0.99\linewidth,bb=163 52 1400 555]{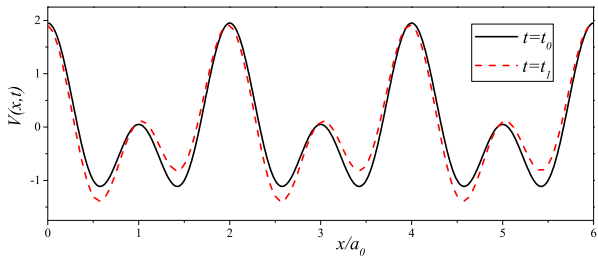}
\caption{(Color online) A time-dependent optical superlattice potential $V(x,t)=V_0\cos(2\pi x/a_0)+0.9V_0\cos[\pi x/a_0-\phi(t)]$ with $\phi(t_0)=0$ and $\phi(t_1)=0.1\pi/a_0$.
} \label{fig:fig7}
\end{figure}

{\it Experimental realization --} The essence of this interactive quantum dynamics can be captured even by a single spin-$\frac 12$ system, which can be realized in most synthetic quantum systems such as superconducting quantum circuit, trapped ions, Rydberg atoms,  as long as their physical quantities can be repeatedly measured during the evolution and Hamiltonian parameters can be dynamically manipulated according to the measurement outcomes. One of the straightforward experimental realization of the two-site problem analyzed above is a 1D ultracold atomic systems with a time-dependent superlattice potential\cite{Nakajima2016}:
\begin{equation}
V(x,t)=V_0\cos(2\pi x/a_0)+0.9V_0\cos[\pi x/a_0-\phi(t)]
\end{equation}
where $a_0$ is the lattice constant, and $\phi(t)$ is phase shift, which will be dynamically modified according to the measurement outcome.  As shown in Fig.\ref{fig:fig7}, such a superlattice system can be approximately considered as a set of disconnected double-well systems (two-site system), in each of them the potential difference $\Delta$  between these two wells can be adjusted by tuning $\phi(t)$. By loading one quantum particle in each double well, and repeatedly measuring the average density difference, one can determine the time-dependence of $\phi(t)$ according to our feedback protocol in Eq.(\ref{eq:chemical}). By performing the average over all the double wells in such a 1D superlattice, one can significantly reduce the measurement-induced uncertainty for the average value of the density.

As for the many-body systems, the proposed localization dynamics is ready to be realized in current ultracold atomic setups in an optical lattice with programmable random disordered potential which can be dynamically modified. As for the detection, the localization results in an initial state memory effect, which can be measured by  preparing a charge-density-wave initial state $|1010\cdots 10\rangle$, then measure the time evolution of density imbalance $I(t)$ via a superlattice band-mapping technique~\cite{Trotzky2012}.

{\it Conclusion and outlook --} In summary, we introduce an interactive quantum dynamics which can give rise to intriguing nonequilibrium steady states.  Future developments will include the generalization of these results to interacting quantum systems, where the interplay between the quantum many-body effect and interactive dynamics may give rise to intriguing non-equilibrium quantum phases of matter. By designing other feedback protocols, it is possible to realize various symmetry breaking phases ({\it e.g.} non-equilibrium superconductor) or even more exotic quantum phases ({\it e.g.} spin liquid) as the steady state of such an interactive quantum system.

{\it Acknowledgments}.---ZC acknowledges helpful discussion with S.Jin and Z.X.Guo. This work is supported by the National Key Research and Development Program of China (Grant No. 2020YFA0309000), NSFC of  China (Grant No.12174251), Natural Science Foundation of Shanghai (Grant No.22ZR142830),  Shanghai Municipal Science and Technology Major Project (Grant No.2019SHZDZX01).


%

\end{document}